\documentstyle[12pt]{article}
\newcommand{\be}{\begin{equation}}
\newcommand{\ee}{\end{equation}}
\newcommand{\bea}{\begin{eqnarray}}
\newcommand{\eea}{\end{eqnarray}}
\newcommand{\nn}{\nonumber}

\begin{document}

\def\theequation{\arabic{section}.\arabic{equation}}
\def\thesection{\Roman{section}}
\def\thesubsection{\Alph{subsection}}
\def\refname{\sc References}

\setcounter{section}0
\setcounter{subsection}0
\setcounter{equation}0

\author{\sf S. A. Gogilidze, \footnotemark[1] ~
A. M. Khvedelidze, \footnotemark[2]~  V. N. Pervushin}

%%%%%%%%%%%%%%%%%%%%%%%%%%%%%%%%%%%%%%%%%%%%%%%%%%%%%%%%%%%%%%%%%%%%%%

\title{\sc Covariant gauges for constrained systems }

%%%%%%%%%%%%%%%%%%%%%%%%%%%%%%%%%%%%%%%%%%%%%%%%%%%%%%%%%%%%%%%%%%%%%
\date{}
\maketitle
\sf
\thispagestyle{empty}

%========================================================
\footnotetext[1]{${}^{*}$%\normalsize
Permanent address: Tbilisi State University,
380086, Tbilisi, Georgia.}
%=======================================================

%====================================================
\footnotetext[2]{%\normalsize
${}^\dagger$
%\normalsize
Permanent address: Tbilisi Mathematical Institute,
380093, Tbilisi, Georgia,\\
\hspace*{0.85cm}Electronic address: khved@theor.jinrc.dubna.su}

%====================================================

\bigskip
\bigskip
\bigskip
\bigskip

The  method of constructing of extended phase space for singular theories
which permits the consideration of
covariant gauges without the introducing of a ghost fields, is proposed.
The extension of the phase
space is carried out by the identification of the initial theory with
an equivalent theory with higher derivatives and applying to it the
Ostrogradsky method of Hamiltonian description .

\bigskip

\centerline{JINR Preprint E2-95-132}
\newpage

\vspace*{2cm}

\setcounter{page}{1}

\section{\sc Introduction}

\bigskip
\bigskip

The crucial point for understanding the physical content of  gauge theories
consist in the identification of field degrees of freedom with  observable
quantities.
The direct  way to achieve this is to handle  only with  physical
degrees of freedom. To realize this program  at  the
classical level one can pass to the theory without  redundant
variables.  This elimination of redundant
variables --- the reduction of initial theory, can be carried out in the
configuration or phase space.  Correspondingly, we have the Lagrangian
or Hamiltonian form of reduction. In the former we restrict the
configuration space to  the consideration of the gauge fields
obeying the Lagrangian constraint equations.  As a result, nonphysical
variables disappear from gauge invariant quantities  and  the remaining
physical  variables describe a usual unconstrained system admitting the
Hamiltonian description.  For realization of the Lagrangian reduction it is
necessary to deal with the explicit solution of constraints.  But it is not
such as a straightforward procedure.  In the first place, in general it
involves a solving of complicated differential equations.  Secondly, apart
>from these practical difficulties there is a desire to maintain the manifest
covariance of the initial gauge theory which is  immediately broken after
resolution of the Lagrangian constraints.  To avoid  these disadvantages
connected with the resolution of constraints in  the Lagrangian reduction
scheme, one commonly uses the  Dirac description for generalized
Hamiltonian systems \cite{DiracL} -  \cite{Unit},
 based on the notion of weak equations.
In this case, the Hamiltonian reduction consists in the elimination of these
week equations. The elimination of weak equations is achieved by
gauge - fixing procedure:  introduction into the theory of
some new ``gauge constraints '' and  replacement of the Poisson bracket by
the Dirac one.  According to the Dirac method, the gauge condition is an
arbitrary function of coordinates and momenta alone. This gauge type of
gauge conditions
has been called the unitary gauges \cite{Unit}.  However, the class of
unitary gauges is not sufficiently large. The so - called covariant gauges
(e.g. the Lorentz gauge  \( \partial_\mu A^\mu = 0 \)) containing the
velocities, which are alien to the Hamiltonian description, turned out it.
So the manifest covariance is again unattainable.  There are known
methods how to overcome  this difficulty.  It has been shown \cite{Yaffe} that
one
can always consider the gauge  with explicit time dependence
which is  equivalent to the velocity depending gauge
 and, thus reduce the problem to the
case of unitary gauges with explicit time dependence.  Another approach has
been established  in ref. \cite{Frad} where with the aim to include this class
of gauges in the consideration the phase space of the initial system is
extended by  introducing  the auxiliary fermionic fields
(ghost fields).

In the  present note we suggest an alternative approach to consideration of
covariant gauges.
In contrast with  the approach \cite{Frad}, the   phase
space for degenerate theory is constructed  without introducing  ghost
fields, by the identification of the initial theory with the equivalent
theory with higher derivatives and applying to it the
Ostrogradsky method of Hamiltonian description \cite{Unit}, \cite{Ostrog}.
We will construct the extended  phase space  ( without ghosts )
and demonstrate that the gauge fixing  condition, which contains a
velocity and thus is nonunitary in ordinary phase space, represents
a usual unitary gauge in the obtained extended  phase space.

The remainder of this article is  organized as  follows.
In the next section, we shall briefly describe the gauge
fixing method according to Dirac's scheme.
Section  3  is  devoted  to  the  Ostrogradsky method of construction
of the phase space for systems with higher derivatives.
(the Legendre transformation generalization)
In  section 4 we describe our scheme.
And in the last section we consider the application of this scheme to the
case of relativistic particle.

%%%%%%%%%%%%%%%%%%%%%%%%%%%%%%%%%%%%%%%%%%%%%%%%%%%%%%%%%%%%%%%%%%%%

\section
{\sc Dirac's ~ method ~of~~ reduction~:~ unitary ~gauge
fixing }

%%%%%%%%%%%%%%%%%%%%%%%%%%%%%%%%%%%%%%%%%%%%%%%%%%%%%%%%%%%%%%%%%%%%

Let us recall briefly the main points of construction of the Dirac
 Hamiltonian description  for degenerate systems .
Suppose, that  the  system with finite number of degrees of freedom
has the following first class constraints  (\(\alpha = 1, \dots, m\))
\be  \label{eq:fcon}
   \phi_\alpha (p,q) \,=\,0, \quad  \{ \phi_\alpha(p,q),
    \phi_\beta (p,q)\} =
f_{ \alpha\beta \gamma} (p,q) \phi_\gamma (p,q).
\ee
The generalized  Hamiltonian dynamics is described by the extended Hamiltonian
\be
H_E =H_C + u_\alpha (t) \phi_\alpha (p, q) \,,
\ee
where the \( H_C \)  is  canonical Hamiltonian  and \( u_\alpha \) are
the Lagrange multipliers .
According to the Dirac  gauge fixing prescription
one must introduce the new  `` gauge '' constraints
\be
\chi_\alpha (p,q) =0 \,                 \label{eq:gauge}
\ee
with  the requirement
\be
\det \{\chi_\alpha, \phi_\beta \} \not= 0,
\ee
 From the  maintenance of these auxiliary conditions (\ref{eq:gauge}) in
time it follows the set of equations
\be
0 \equiv \dot{\chi}_\alpha = \{\chi_\alpha, H_C \} +
\sum_\beta \{\chi_\alpha,
\phi_\beta  \} u_\beta = 0,
\ee
which allows determining the unknown Lagrange multipliers.
Formally, the solution can be written as
\be \label{eq:lm}
u_\beta =  - \sum \{H_C, \chi_\alpha \}C^{-1}_{\alpha \beta}\phi_\beta \,,
\ee
where  \(C^{-1} \) is the inverse  matrix of
\[
C_{\alpha \beta} = \{\chi_\alpha, \phi_\beta\},   \quad
C_{\alpha \beta}  C^{-1}_{\beta \gamma} = \delta _{\alpha \gamma}
\]
The Dirac elegant observation consists in  that instead of the
determination in such a way the unknown function
\(u_\alpha(t)\),  one can change  the Poisson brackets to the  new ---
Dirac one
\be
\{F, G \}_{D} = \{F, G \} -
\{F \chi _\alpha\}C^{-1}_{\alpha \beta} \{ \phi_\beta, G \},
\ee
Since, all constraints, including gauge one, have zero Dirac's brackets
with everything, we can consider them as strong equations.
This change  of brackets  reflects the reduction in
number of  degrees of freedom. It is easy verify  that
\[
\sum_{i =1}^{n}
\{q_i, p_i \}_{P.B.} = n \quad
\sum_{i =1}^{n}
\{q_i, p_i \}_{D}
= {n-m}
\]
So the Dirac brackets take into account the constrained character of
the  theory and effectivelly reduce the
phase space.

As it has been mentioned in the introduction, this  method
works only for gauges, known as unitary gauges \cite{Unit}.
To overcome this restriction one can proceed  as follows.
 Suppose, the gauge condition depends on
derivatives of the \(r+l\) - th order of coordinates with respect to the time
\be \label{eq:covgauge}
\chi_\alpha (p, q^{(0)}, q^{(1)}, q^{(2)},
\dots ,  q^{(r+l)}) = 0 \,, \quad q^{(k)} \equiv
\frac{d^{k}}{dt^{k}} q(t)
\ee
The main idea is to consider the enlarged
configuration space by rising these  higher order derivatives the status of
coordinates.
This can be achieved by  identifyning  the  initial theory with equivalent
theory with higher derivatives and applying to it the
Ostrogradsky method for construction of the phase space.
More precisely, suppose that \(r \) - is the maximum of the order of
derivative with respect to the time  \(
q^{(r)} \) entering in to the Lagrangian \({}^{(r)}L (q, q^{(1)},
\dots, q^{(r)} ) \) while the highest order of time derivative in
gauge fixing condition is \(r+l \).Then, instead of do the Legendre
transformation
on the coordinates of the configuration  space of the initial Lagrangian
\(L(q,\dots ,{q}^{(r)} )\) we suggest dealing  with the Legendre
transformation for the new Lagrangian \({}^{(r+l)}L^{\ast} \) defined as a
function of fictional auxiliary variables \(q^{(r+1)}, \dots, q^{(r+l1)} \)
by the anzatz
\[
{}^{(r+l)}L^{\ast}(q, q^{(1)}, \dots, q^{(r+l)}  ) = {}^{(r)}L(
q, q^{(1)}, \dots, q^{(r)}  )
\]
 The phase obtained for the new equivalent theory will be an  extended
phase space for the initial system.
And our gauge  fixing  conditions
(\ref{eq:covgauge}),  nonunitary in the ordinary phase space, now becomes
the usual unitary gauge in the extended phase space.  To
prove this, let us briefly describe the Ostrogradsky method  of
the Hamiltonian formulation of theories with higher derivatives
\cite{Unit}, \cite{Ostrog}.

%%%%%%%%%%%%%%%%%%%%%%%%%%%%%%%%%%%%%%%%%%%%%%%%%%%%%%%%%%%%%%%%%%%%%%%%%%
\section{\sc Ostrogradsky method for theories with higher derivatives}
%%%%%%%%%%%%%%%%%%%%%%%%%%%%%%%%%%%%%%%%%%%%%%%%%%%%%%%%%%%%%%%%%%%%%%%%%%%%

Let us consider the variational problem for the unconstrained mechanical system
\footnote{\normalsize We assume that q is the \( n \) - dimensional vector}
\[
S[ q ] =\int dt L  ,
\]
descrbed by the Lagrangian  \( L(q, q^{(1)}, q^{(2)},\dots ,  q^{(k)})
\) which is a function not only of coordinate \( q \) and velocity
\(\dot{q}\) but also of  higher derivatives of coordinate with respect
to the time.  The Euler - Lagrange equation follows from the extremum condition
\( \delta S = 0 \)
\be  \label{eq:EL}
\sum_{s = 0}^{k} (-1)^{s} \frac{d^{s}}{dt^{s}} \left( \frac{\partial L}
{\partial q^{(s)}} \right) = 0
\ee
with the zero boundary conditions for variations
\( \delta q^{(i)} \vert_{boundary} = 0 \;\; i = 0, \dots, k-1 \).
Thus, in general the Euler -  Lagrange equation has order \( 2k \).
The Ostrogradsky theorem stands that there is a generalized
Legendre transformation such that this Euler - Lagrange equations
for the non - singular \( n \) - dimensional system (\ref{eq:EL}) can be
transformed into the equivalent set of first order equations of
the Hamiltonian form on a space of dimension \( 2kn \).
To construct the Legendre transformation, let us introduce
\(2kn \)
canonical variables \( \xi_i \) and \( p_i
\;\; i = 1, \dots, k\)
\bea \label{eq:phase}
  \xi_i& = & q^{(i-1)},  \\
  p_i& = &\sum_{s =i}^{k} (-1)^{s-i} \frac{d^{s-i}}{dt^{s-i}} \left(
 \frac{\partial L}{\partial q^{(s)}} \right)
\eea
and then set the function
\be
H(\xi,\dot{\xi}) = - L + \sum_{i = 1}^{k-1} p_i \xi_{i+1} + p_k \dot{\xi}_k
\ee
where the Lagrangian \( L \) in terms of new variables  has the form
\[
 L(q, q^{(1)}, q^{(2)},\dots, q^{(k)})
= L(\xi_1, \xi_2,\dots ,\xi_k, \dot{\xi_k})
\]
Now, taking into account that
for nonsingular Lagrangian we can express from
(\ref{eq:phase}) the velocity \(\dot{\xi_k}\) as a
function of the remaining r variables
\[ \dot{\xi_k} = f(\xi, p ),
\]
one can, according to  Ostrogradsky's theorem \cite{Unit}, \cite{Ostrog},
rewrite the Euler - Lagrange equations (\ref{eq:EL}) in the Hamiltonian form
\bea
\label{eq:heq}
\dot{\xi_i} & =& \{ \xi_i, \overline{H} \}, \nn\\
\dot{p_i}  & = &\{ p_i, \overline{H}\},
\eea
with the Hamiltonian \(\overline{H} (\xi,
p)\) defined by
\bea
\overline{H}(\xi, p) = {H} \vert_{\dot{\xi_k} = f(\xi, p )},
\eea
and the Poisson brackets in (\ref{eq:heq}) have the  form
\[
\{F,G \} =\sum_{s=1}^{k} \left ( \frac{\partial F}{\partial \xi_{s}}
\frac{\partial G}{\partial p_{s}}
 - \frac{\partial F}{\partial p_
{s}}\frac{\partial G}{\partial \xi_{s}}
\right )
\]

%%%%%%%%%%%%%%%%%%%%%%%%%%%%%%%%%%%%%%%%%%%%%%%%

\section{\sc Enlarged phase space for covariant gauges}

%%%%%%%%%%%%%%%%%%%%%%%%%%%%%%%%%%%%%%%%%%%%%%%%%%%%%

Now we are ready to return to the case of the singular theory with
the first class constraints (\ref{eq:fcon}) and to consider
the  covariant gauges of the
type  (\ref{eq:covgauge})
\[
\chi_\alpha (p, q, q^{(1)}, q^{(2)},
\dots ,  q^{(l+1)}) = 0
\]
Let us
correlate with  the initial singular theory, describing by Lagrangian
\(
L ( q,{\dot q})
\)
the { \underline{ equivalent}} theory for which the new
Lagrangian is considered as a function of additional derivatives of
the coordinate with respect to the
time variable up to the l-th order ( \(l \) is a maximal order of a time
derivative in the gauge condition ( \ref{eq:covgauge}))
\be
L^{\ast} ( { q},{q}^{(1)},\dots, {q}^{(l)}) =
L({q}, \dot{q})
\label{eq:NL}
\ee
The  method of construction
of phase space picture for non singular Lagrangians
described in previous section
admit the generalization to the  singular case in Dirac's sense
\cite{Unit}.
Application of the above written procedure
to new Lagrangian \( L^{\ast}\)  gives  the following.
According to the definition of canonical conjugate variables
(\ref{eq:phase}), for the theory (\ref{eq:NL})  we obtain
the usual part of canonical variables \(q, p \)
\bea
    {\xi}_1 & = & q^{(0)} = q, \nn \\
    {p}_1   & = &  \frac{\partial L}{\partial q^{1}} = p ,
\eea
and additional one
\({\xi}^{\ast}_i,  {p}^{\ast}_i, \quad i = 2, \dots, l+1 \)
\bea
    {\xi}^{\ast}_i & = & q^{(i-1)},  \nn\\
    {p}^{\ast}_i & = &
\sum_{s =i}^{l+1} (-1)^{s-i} \frac{d^{s-i}}{dt^{s-i}} \left(
 \frac{\partial L^{\ast}}{\partial q^{(s)}} \right)  ,
 \label{eq:1phase}
\eea
 From (\ref{eq:1phase}) it follows that
in addition to the  old constraints
(\ref{eq:fcon}) we obtain  a new set of \( l \) constraints
\bea \label{eq:1newconst}
    {p}^{\ast}_i & = & 0, \quad i = 2, \dots l+1.
\eea
Taking into account that the new canonical Hamiltonian coincides with the
initial one
\[
H_C^{\ast} (\xi^{\ast} p^{\ast}) = H_C (\xi_1, p_1 )
\]
one can verify that there aren't  new secondary constraints,
and we obtain the
extended  Hamiltonian  for the new theory  in terms of the old one
\(H_{E} (\xi_1, p_1 )\)
\be
H_{E}^{\ast} (\xi^{\ast} p^{\star}) = H_{E} (\xi_1, p_1 ) +
\sum_{s=2}^{l+1} u_s^{\ast}p^{\ast}_s ,
\ee
with  new Lagrangian multipliers \( u^\ast \).
So in the extended phase space \((
\xi^{\ast} p^{\ast}, \xi_1, p_1 ) \)
we get the generalized dynamics with
the set of constraints
\bea \label{eq:totconst}
\phi_\alpha (p,q) & = & 0, \quad  \alpha = 1, \dots, m,\nn\\
    {p}^{\ast}_i & = &  0, \quad i = 2, \dots, l+1,
\eea
with the desired condition (\ref{eq:covgauge}), that
has the form of a unitary gauge depending only on canonical variables
\be
\chi_\alpha (\xi^\ast , p^\ast, \xi_1, p_1) = 0 .
\ee
Now if we  wants to reduce the phase space of our theory
 due to the appearance of new
constraints (\ref{eq:1newconst}), we need  additional \( l \) -
gauge conditions
\[
\chi^{\star}_\alpha (\xi_1, p_1,  p^\ast, \xi^\ast) = 0 \, ,\quad
\alpha = 2,\dots , l+1 .
\]

%%%%%%%%%%%%%%%%%%%%%%%%%%%%%%%%%%%%%%%%%%%%%%%%%%%%%%%%%

\section{\sc Gauge ${x}_\mu \dot{x}^\mu = 0$
for a relativistic particle}

%%%%%%%%%%%%%%%%%%%%%%%%%%%%%%%%%%%%%%%%%%%%%%%%%%%%%%%%

In this section, we will apply  the above described scheme  to
the simple example, the  relativistic particle with mass \(m\)
\be\label{eq:partaction}
W[x] = - m\int\limits_0^{T} d\tau \sqrt{\dot x_\mu^2} .
\ee
This action is invariant under reparametrization of time
\bea
\tau \to \tau'& = &s(\tau),\nn\\
x(\tau) \to x(\tau')' & = &  x(\tau),
\eea
with $ds/d\tau >0$. Therefore, there is an arbitrary function in the
solution for the equation of motion .
As a consequence of this invariance, we have  identically vanishing
canonical Hamiltonian and energy shell constraint
\be  \label{eq:es}
\varphi^1 \equiv p^2 - m^2  = 0.
\ee
Thus, the total Hamiltonian is
\[
H_T = u(t) (p^2 - m^2).
\]
There is  some freedom in the definition of dynamics in the reparametrized
invariant theory. This fact is reflected in the gauge fixing
procedure.
It is known that in this case the gauge fixing
condition with  necessity explicitly  depends on the time
parameter. For example, the proper time fixing
\[
x_0(\tau) = \tau
\]
corresponds to the instant form of dynamics for relativistic particle .
In this gauge, one can according to (\ref{eq:lm}), fix the function
\(u(t)\)
\be
u(t) = \frac{1}{2\sqrt{p_i^2+ m^2}}
\ee
and get the  usual equations of motion
\be
\dot{x}_i = \frac{p_i}{\sqrt{p_i^2+ m^2}}
\ee
governed by the reduced Hamiltonian \( H_{Red} = \sqrt{p_i^2+ m^2} \)
\cite{Hanson}.
How cfn one can get this results
in a gauge with the dependence onthe  velocity \({\dot{x}^\mu} \),
for example, of a following type:
\be
\xi \equiv \dot{x}^\mu {x}_\mu = 0
\ee
To do this, we  will  act in the spirit of the previous section.
Let us pass from the initial singular theory, described  by
the action (\ref{eq:partaction})
to the equivalent theory for which the new
Lagrangian is considered as a function of a first and second order
derivatives of coordinates with respect to the time.
After the transformation to the enlarged  configuration space
with the coordinates
\[
\xi_\mu = x_\mu , \quad  \xi^{\ast}_\mu = \dot{x}_\mu  ,
\]
\be \label{eq:newL}
L^{\ast} ( \xi, \xi^{\ast}, \dot{\xi}^{\ast} ) =
- m \int\limits_0^{T} d\tau \sqrt{ {\xi^{\ast}}_\mu^2},
\ee
we see that in addition to  the
old constraints (\ref{eq:es}) there are  new ones
\be  \label{eq:1newconstr}
p_\mu^{\ast} = \frac{\partial L^{\ast}}{\partial \xi^\ast_\mu} = 0 .
\ee
Taking into account that
new secondary constraints do not arise we obtain the
extended  Hamiltonian for the new theory
\be
H_{E}^{\ast}  =   u(t)(p^2 - m^2)
+ u_\nu^{\ast}p^{\ast}_\nu ,
\ee
with the new Lagrangian multipliers \(u^\ast_\nu \).
The gauge condition \(\dot{x}^\mu {x}_\mu = 0 \) in the enlarged phase space
has  now the form of unitary gauge
\be
\chi = \xi^{\ast}_\mu \xi^{\mu} = 0
\ee
Let us choose  additional gauge conditions
for elimination of new constraints
(\ref{eq:1newconstr})
\[
\chi^{\star}_\nu \equiv \xi^\ast_\nu - a_\nu(\tau) = 0
\]
with  some  functions \(a_\nu(\tau) \).
With  the help of these gauge conditions
one can find the Lagrange multipliers and verify that the
choice of this vector leads to one or another form of dynamics.
The choice $a_\mu(\tau) = (a(\tau), 0, 0, 0)$  corresponds
to the
instant form of dynamics  with the time variable $x_0 = a(\tau) $
while for the  light - like vector $a_\mu(\tau) = (a(\tau), 0, 0, a(\tau))$  we
get the
light cone formulation for the  relativistic particle dynamics with the
light cone time \( x_{+}\equiv x_0 + x_3 = 2 a(t) \).

The authors  would like to thank A.N. Kvinikhidze,  V.V.  Nesterenko and
Yu.S. Surovtsev for helpful comments and discussions.

\end{document}